\RequirePackage{fix-cm}
\documentclass[twocolumn,epjc3]{svjour3}  
\smartqed  
\RequirePackage{graphicx}
\journalname{Eur. Phys. J. C}
\usepackage{lipsum}
\usepackage{lineno,hyperref}
\usepackage{lineno}
\usepackage{graphicx}
\usepackage{subcaption}
\usepackage{amsmath}
\usepackage{verbatim}
\usepackage{color}
\usepackage{soul}
\usepackage[dvipsnames]{xcolor}
\usepackage{mathrsfs}
\modulolinenumbers[5]

\begin{document}

\title{Shape-invariant Potentials and Singular Spaces
}


\author{Peng Yu\thanksref{addr1}
        \and
        Yuan Zhong\thanksref{e1,addr1} 
        \and
       Hui Wang\thanksref{addr1}
        \and
        Ziqi Wang\thanksref{addr1}
        \and
        Mengyang Zhang\thanksref{addr2}
}

\thankstext{e1}{e-mail: zhongy@mail.xjtu.edu.cn {(corresponding author)}}


\institute{MOE Key Laboratory for Nonequilibrium Synthesis and Modulation of Condensed Matter, School of Physics, Xi'an Jiaotong University, Xi'an 710049, China \label{addr1}
           \and
           Institute for Quantum Science and Technology (IQST) and Department of Physics, Shanghai University, Shanghai 200444, China \label{addr2}}

\date{Received: date / Accepted: date}

\maketitle

\begin{abstract}
In this work, we present two brane-world-type solutions in a two-dimensional (2D) dilaton gravity model with singular space-time backgrounds. By employing a first-order superpotential formalism, we first construct the 2D analogues of the thick brane solution previously given by Gremm and analyze the corresponding linear scalar perturbations. We show that for a model with canonical scalar matter fields,  the effective potential of the linear perturbation equation is a singular P\"oschl--Teller~II type, which does not admit bound states. However, for a model with non-canonical scalar fields, the effective potential becomes an exactly solvable P\"oschl--Teller~I potential, which has an infinite tower of normalizable bound states. We also present a second analytic solution inspired by the work of Girardello \emph{et al.}, but with non-canonical scalar field. In this case, the linear perturbation equation is a Schr\"odinger equation with the Eckart potential, which is also exactly solvable.
\keywords{2D brane world solutions \and P\"oschl-Teller I potential \and Eckart potential \and 2D gravity}
\end{abstract}






\section{Introduction}\label{introduction}
In an early study, Gremm constructed a five-dimensional (5D) thick brane solution in a gravity-scalar field system, where the brane configuration interpolates between regions with naked singularities~\cite{64}, in contrast to earlier works that primarily considered branes connecting spaces with AdS {geometries}~\cite{DeWolfe:1999cp,Gremm:1999pj,55,54}. {{In Gremm's work, the metric takes the following form 
\begin{equation}
ds^2=e^{2 A(r)} \eta_{\mu \nu} d x^{\mu} d x^{\nu}+d r^{2}, \nonumber
\end{equation}
where the warp factor
\begin{equation}
A(r)=n \log [\cos (k r)] \label{003}
\end{equation}
is singular at $r=\pm \frac{\pi}{2 k}$.}} Usually such singular spaces should be discarded, but Gremm argued~\cite{64} that such singularities may admit a physical interpretation as duals of strongly coupled, non-conformal field theories, according to the theory of AdS/CFT correspondence~\cite{98}. 

An important result from Gremm's work is that the tensor perturbation spectrum has two normalizable bound states and a continuum that is asymptotic to plane waves, with a mass gap $m_{\mathrm{gap}} = \tfrac{3k}{2}$. For a four-dimensional (4D) brane observer, a normalizable zero mode reproduces the Newtonian potential, while continuum modes generate corrections $\Delta U(r) \sim \frac{1}{r^{2}}$, 
see~\cite{55}. {{Because the continuum modes can carry flux into singularities, and spoil 4D energy conservation, one usually imposes unitary (no-flux) boundary conditions to eliminate the continuum states and keeping only the discrete states, to obtain a closed 4D effective theory. However, from the perspective of  AdS/CFT theory, such boundary flux of the continuum states are permitted~\cite{64}, as they can be interpreted as energy transfer to a dual $\mathrm{IR}$ sector. }}

With the tensor sector investigated, the next step is the scalar sector, which is not treated by Gremm but is essential for stability and holographic interpretation~\cite{98}. However, scalar perturbations in 5D brane-world backgrounds are more intricate than the tensor sector, analytic progress is often limited. To retain analytic control while preserving the essential physics, it is useful to turn to 2D models. 2D models are simple and powerful tools for addressing diverse problems, including quantization of gravity~\cite{1,2}, black hole physics~\cite{3,4,5}, cosmology~\cite{7,8,9} and relativistic gravitational N-body problem~\cite{Mann:2024syg}. But in 2D space-time, the Einstein tensor always vanishes for any metric, so one usually introduces a dilaton field  to generate non-trivial gravitational dynamics~\cite{89,90}.

A well-known 2D gravity model that resembles Einstein gravity in many aspects, is the so-called Mann-Morsink-Sikkema-Steele (MMSS) model~\cite{17}, whose dynamical equations have a simple first-order formalism, from which exact gravitating kink solutions can be easily constructed~\cite{56,53,51,38,39,84,85,86,Lima:2022chw,Andrade2022}. Many of these solutions have asymptotic $\rm AdS_{2}$ {{{metrics}}}, and can be regarded as 2D versions of some 5D thick brane solutions~\cite{DeWolfeFreedmanGubserKarch2000,Gremm2000,CsakiErlichHollowoodShirman2000,30,88}.  
Moreover, it has been shown that, if one introduces {non-canonical scalar matter fields to this model, it is possible to obtain gravitating-kink solutions with exactly solvable linear stability potentials} such as P\"oschl-Teller~II~\cite{38} and Rosen-Morse~\cite{39}. However, the works in Refs.~\cite{38,39} focused on smooth space-time backgrounds free of singularities. It is thus natural to ask whether {{kink solutions with exactly solvable linear stability potentials also appear in singular space-times}}. 

In this work, we first construct the 2D analogue of Gremm's solution in the MMSS model and analyze scalar perturbations for both canonical and {{non-canonical}} scalar matter fields. We find that for both cases, the effective potentials in the Schr\"odinger equations from linear perturbations are exactly solvable. However, only the effective potential in the {{non-canonical}} model supports bound states. Unlike Gremm's 5D construction, where analytic forms of the effective potentials from scalar perturbations exist only for the special cases ($n=1,2$), our 2D {{non-canonical}} model admits closed analytic expressions of the effective potential for arbitrary values of the parameter $n$. {{Then,}} inspired by the works on holographic RG flow of Girardello, Petrini, Porrati and Zaffaroni (the GPPZ flow)~\cite{Girardello:1999bd,63}, we construct another analytic solution in our 2D model, where linear perturbations yield another exactly solvable Eckart potential.  

The paper is organized as follows. In Sec.~\ref{sec:Model} we give the model and general analysis of scalar perturbations. In Sec.~\ref{examples}, we first construct a 2D analogue of Gremm's solution via a first-order formalism and then analyze its linear scalar perturbations. We then present another analytic solution whose linear perturbations lead to the Eckart potential without bound states. We conclude with a summary of the main results of this work in Sec.~\ref{sec:conclusion}.

\section{The model and linear perturbations}
\label{sec:Model}
We work in the framework of Mann--Morsink--Sikkema--Steele (MMSS) model given by the action~\cite{17}:
\begin{equation}
\label{05}
S=\frac{1}{\kappa} \int d^2x\,\sqrt{-g}\left[-\frac{1}{2}(\nabla \varphi)^2+\varphi R+\kappa \mathcal{L}(\phi,X)\right],
\end{equation}
{{where $\kappa$ is the 2D gravitational coupling constant, $R$ is the Ricci scalar,}} $\varphi$ is the dilaton, $\mathcal{L}(\phi,X)$ is the matter Lagrangian, taken here to be a real scalar field and $X=-\frac{1}{2} \nabla^\mu \phi \nabla_\mu \phi$ is the kinetic term of the scalar matter Lagrangian. For the static metric ansatz:
\begin{equation}\label{1}
ds^2=-e^{2A(r)} dt^2+dr^2,   
\end{equation}
the equations admit a first-order formalism that allows one to construct gravitating kink solutions systematically~\cite{56,53,51,38,39,84,85,86}.

After variations of Eq.~(\ref{05}), one obtains the Einstein equation, the scalar {{equation,}} and the dilaton equation. For the metric in Eq.~(\ref{1}), one finds that the dilaton field is related to the warp factor $A(r)$ by:
\begin{equation}\label{01}
\varphi(r)=2 A(r).
\end{equation}
{{Inserting}} the above equation into the Einstein equation, one obtains the following two equations:
\begin{eqnarray}
-4 \partial_r^2 A=\kappa \mathcal{L}_X\left(\partial_r \phi\right)^2 &,& \label{091}\\
4 \partial_r^2 A + 2\left(\partial_r A\right)^2 = \kappa \mathcal{L}&,&  \label{010} 
\end{eqnarray}
where $\mathcal{L}_X \equiv \frac{\partial \mathcal{L}}{\partial X}$. 
\begin{figure*}[!ht]
  \centering 
  \includegraphics[width=0.5\textwidth, angle=0]{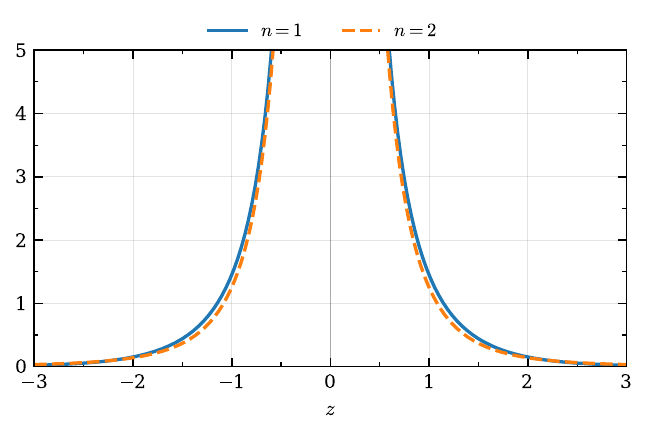}	
  \caption{Plots of the effective potentials $V(z)$ for scalar perturbations in the 2D canonical model with $n=1,2$.} 
  \label{fig:4}%
\end{figure*} 
In the canonical model where $\mathcal{L}=X-V(\phi)$, the dynamical equations (\ref{091}) and (\ref{010}) can be written as a very simple first-order formalism:
\begin{eqnarray}
  \partial_r A=-\frac{1}{4} \kappa W, \label{23}\\
  V=\frac{1}{2} W_\phi^2-\frac{1}{8} \kappa W^2,\label{24}
\end{eqnarray}
if we introduce the so-called superpotential $W(\phi)$ by demanding{ 
\begin{equation}\label{22}
  \partial_r \phi=W_\phi \equiv\frac{d W}{d \phi},
\end{equation}
which also equivalents to
\begin{equation}\label{011}
X=-\frac{1}{2} W_\phi^2
\end{equation}
for static solutions.
}

The advantage of this formalism is that, by carefully choosing the superpotential $W(\phi)$, one can easily obtain exact kink solutions as shown in~\cite{DeWolfeFreedmanGubserKarch2000,Gremm2000,CsakiErlichHollowoodShirman2000} and~\cite{53}. 

For {{non-canonical}} models where $\mathcal{L} \neq X-V(\phi)$, the Einstein equations usually take more complicate forms than Eqs.~(\ref{23}) and (\ref{24}). However, for the following Lagrangian:
\begin{equation}\label{25}
  \mathcal{L}=U(\phi)\left(X+\frac{1}{2} W_\phi^2\right)^2+X-V(\phi), 
\end{equation}
$\mathcal{L}$ and $\mathcal{L}_X$ will be the same form as those of the canonical case, if Eq.~(\ref{011}) is satisfied~\cite{38,39}. Thus, the dynamical equations derived from the above Lagrangian retain the same structure as in the canonical case, leading to the same background solutions. {{However, differences appear}} as soon as linear perturbations are considered, as will be shown in the following. {{The model described by Eq.}}~(\ref{25}) {{is called a twinlike model of the canonical model. The twinlike models are distinct field theories that share identical background solutions and energy densities, but may differ in their perturbative properties. They are first discussed in cosmology and brane world models}}~\cite{95,96,97}.

Now we consider linear scalar perturbations. To perform the analysis, we first introduce a coordinate transformation:
\begin{equation}
z \equiv \int e^{-A(r)} d r,    
\end{equation}
with which the metric in Eq.~(\ref{1}) becomes conformally flat:
\begin{equation}
d s^2=e^{2 A(z)}\left(-d t^2+d z^2\right).    
\end{equation}
Suppose we have a static background solution \{$\varphi(z)$, $\phi(z)$, $A(z)$\} and let \{$\delta \varphi(z,t)$, $\delta \phi(z,t)$, $\delta g_{\mu \nu}(z,t)$\} be small perturbations. For convenience, we define~\cite{51,38}:
\begin{eqnarray}
\delta g_{\mu \nu}(z, t) &\equiv& e^{2 A(z)} h_{\mu \nu}(z, t) \nonumber \\
&=&e^{2 A(z)}\left(\begin{array}{cc}
h_{00}(z, t) & \Phi(z, t) \\
\Phi(z, t) & h_{z z}(z, t)
\end{array}\right),
\end{eqnarray}
and $\Xi \equiv 2 \dot{\Phi}-h_{00}^{\prime}$ (where overdots and prime denote derivatives with respect to time and space, respectively) and work with the gauge condition $\delta \varphi(z)=0$. After linearization we obtain three independent perturbation equations. Two of them can be used to eliminate $\Xi$ and $h_{zz}$ in terms of $\delta \phi$, and finally one can obtain an equation of the following form~\cite{51}:
\begin{equation}
  -\partial_t^2 G+\partial_y^2 G-V_{\rm {eff }}(y) G=0   , 
  \end{equation}
after performing another coordinate transformation $z \rightarrow y=\int d z \gamma^{-1 / 2}$. Here, $G$ and $\gamma$ {{are}} defined as:
\begin{eqnarray}
  G(y, t) &\equiv& \mathcal{L}_X^{1 / 2} \gamma^{1 / 4} \delta \phi(y, t) ,\\
  \gamma &\equiv& 1+2 \frac{\mathcal{L}_{X X} X}{\mathcal{L}_X} ,
\end{eqnarray}
{{where $\mathcal{L}_{X X} \equiv \frac{\partial^2 \mathcal{L}}{\partial X^2}$. The effective potential}} is:
\begin{equation}\label{34}
  V_{\mathrm{eff}}(y) \equiv \frac{\partial_y^2 f}{f} ,
\end{equation}
where
\begin{equation}\label{35}
  f(y) \equiv \mathcal{L}_X^{1 / 2} \gamma^{1 / 4} \frac{\partial_y \phi}{\partial_y A} . 
\end{equation}
Obviously, to make the rescaling of $\delta \phi$ as well as the coordinate transformation $z \to y$ well defined, the following conditions must be satisfied~\cite{51}:
\begin{equation}
\mathcal{L}_X>0, \quad \gamma>0    . 
\end{equation}
By performing {{a mode expansion}}:
\begin{equation}
G(t, y)=\sum_l e^{i \omega_l t} \psi_l(y),    
\end{equation}
we obtain a Schr\"odinger-like equation with factorizable Hamiltonian:
\begin{equation}
H \psi_l \equiv \mathcal{A} \mathcal{A}^{\dagger} \psi_l=-\frac{d^2 \psi_l}{d y^2}+V_{\rm eff}(y) \psi_l=\omega_l^2 \psi_l,
\end{equation}
where,
\begin{equation}
\mathcal{A}=\frac{d}{d y}+\mathcal{W}, \quad \mathcal{A}^{\dagger}=-\frac{d}{d y}+\mathcal{W}, 
\end{equation}
and,
\begin{equation}\label{40}
  \mathcal{W} \equiv \frac{\partial_y f}{f}.
\end{equation} 
The $\mathcal{W}$ above is also called the superpotential in the context of supersymmetric quantum mechanics (SUSY QM), but is different from the one defined in Eqs.~(\ref{22}) and (\ref{011}). In SUSY QM, one can construct two partner Hamiltonian operators:
\begin{equation}
H_{ \pm}=-\frac{d^2}{d y^2}+V_{ \pm}(y),
\end{equation}
where $V_{ \pm}(y)=\mathcal{W}^2 \pm \frac{d \mathcal{W}}{d y}$ are called partner potentials. If the partner potentials are non-singular, the eigenvalues of the corresponding partner Hamiltonians are non-negative, and have equal spectra except the ground state \cite{22}. Moreover, the partner potentials are called {{shape-invariant}} if they satisfy the following equation:
\begin{equation}
V_{+}\left(y ; a_0\right)=V_{-}\left(y ; a_1\right)+F\left(a_1\right),  
\end{equation} 
where $a_0$ and $a_1$ are two constants and $F(a_1)$ is an arbitrary function { of $a_1$}~\cite{22}. In this case, the corresponding Schr\"odinger equation is exactly solvable \cite{24,25}.

Note that to derive the Schr\"odinger-like equation, one needs to introduce two coordinate transformations $r \to z \to y$. The above Schr\"odinger-like equation is obtained in the $y$-coordinates while background solutions are usually derived in the $r$-coordinates. Usually, it is hard to get the analytical expression of the effective potential $V_{\rm eff}(y)$, needless to say to solve the stability equation exactly. However, it was found that for some models with {{non-canonical}} scalar Lagrangian, such as the one in Eq.~(\ref{25}), the second coordinate transformation inverses the first one, provided that~\cite{38}:
\begin{equation}\label{06}
  U(\phi)=\frac{1}{2 W_\phi^2}\left(1-e^{\frac{\kappa}{2} \int \frac{W}{W_\phi} d \phi}\right).
  \end{equation}
This option makes the $y$-coordinate system and the $r$-coordinate equivalent and in this case, it is possible to obtain both the background solutions and the Schr\"odinger-like equations in the $r$-coordinate. And in general, given the superpotential $W(\phi)$, we can use Eqs.~(\ref{22}), (\ref{23}) and (\ref{34}) to find $\phi(r)$, $A(r)$ and $V_{\rm eff}(r)$, respectively. To show this, we will present two explicit examples in the following and give scalar perturbation analysis. 

\section{Explicit examples}\label{examples}
\subsection{2D analogues of Gremm's solution}
It has been shown in Ref.~\cite{60} that Gremm's 5D thick brane solution can be produced by the superpotential:
\begin{equation}\label{008}
W(\phi)=k v^{2}\sinh\!\left(\frac{\phi}{v}\right),
\end{equation}
from a first-order formalism. If we take the above superpotential in our 2D model, we obtain a 2D analogue of Gremm's solution from the first-order formalism discussed in Sec.~\ref{sec:Model}. The solution reads: 
\begin{eqnarray}
\phi(r)&=&v\,{\rm arcsinh}\!\big(\tan(k r)\big),\label{48}\\
A(r)&=&\tfrac{1}{4}\,\kappa v^{2}\,\log\!\big(\cos(k r)\big),\label{0149}\\
V(\phi)&=&-\tfrac{1}{8}\,k^2 v^{2}\!\left(\kappa v^{2}\sinh^{2}\!\frac{\phi}{v}-4\cosh^{2}\!\frac{\phi}{v}\right)\label{050},
\end{eqnarray}
{ where $k$ and $v$ parametrize the thickness of the solution and the expectation value of the vacua, respectively. As expected, the warp factor diverges at $r=\pm \frac{\pi}{2 k}$, the boundaries of the space-time.} Note that here the coefficient $\kappa v^{2}/4$ takes the role $n$ as compared to Eq.~(\ref{003}), so in the following, we denote $\kappa v^{2}/4=n$ and further set $k=1$. 

Since Gremm's model is canonical, we first analyze scalar perturbations in the canonical model before turning to the {{non-canonical}} case.

\subsubsection{Scalar perturbations in the canonical model}
In the canonical model, the Lagrangian is $\mathcal{L}=X-V(\phi)$, so that $\mathcal{L}_X=1${, $\gamma=1$ and $y=z$.} As noted in Ref.~\cite{64}, the coordinate transformation $r\to z$ is analytic only for $n=1,2$.

For the first analytic case where $n=1$, the coordinate transformation is given by $z=\ln \left(\frac{1+\tan (r / 2)}{1-\tan ( r / 2)}\right)$, and the background solutions are:
\begin{eqnarray}
\phi(z)&=&v\,z,\\
A(z)&=&-\log(\cosh(z)).
\end{eqnarray}
Note that { although $r$ is restricted to the domain $(-\frac{\pi}{2},\frac{\pi}{2})$,  $z$ can run from $-\infty$ to $+\infty$}. 

From Eqs.~{\eqref{34} and \eqref{35}}, the effective potential is:
\begin{equation}\label{0012}
V(z)=2\,\mathrm{csch}^{2}(z), \quad \text{for} \quad  n=1.
\end{equation}
{{This is the singular P\"oschl-Teller II potential:}}
\begin{align}
  V(z ; B, C) &= (B-C)^2 - B(B-1) \operatorname{sech}^2({{z}}) \nonumber\\
  &\quad + C(C+1) \operatorname{csch}^2({{z}}),
\end{align}
{{with $B=1$}} and $C=1$, appeared previously in the work~\cite{38}, and is known to be shape-invariant in {{SUSY QM}}. As can be seen in Fig.~\ref{fig:4}, $V(z)\ge0$ everywhere and decays asymptotically as $V(z)\sim 2e^{-2|z|}$ for $|z|\to\infty$, there will be no bound states and the solutions are stable.
\begin{figure*}
	\centering 
	\includegraphics[width=\textwidth]{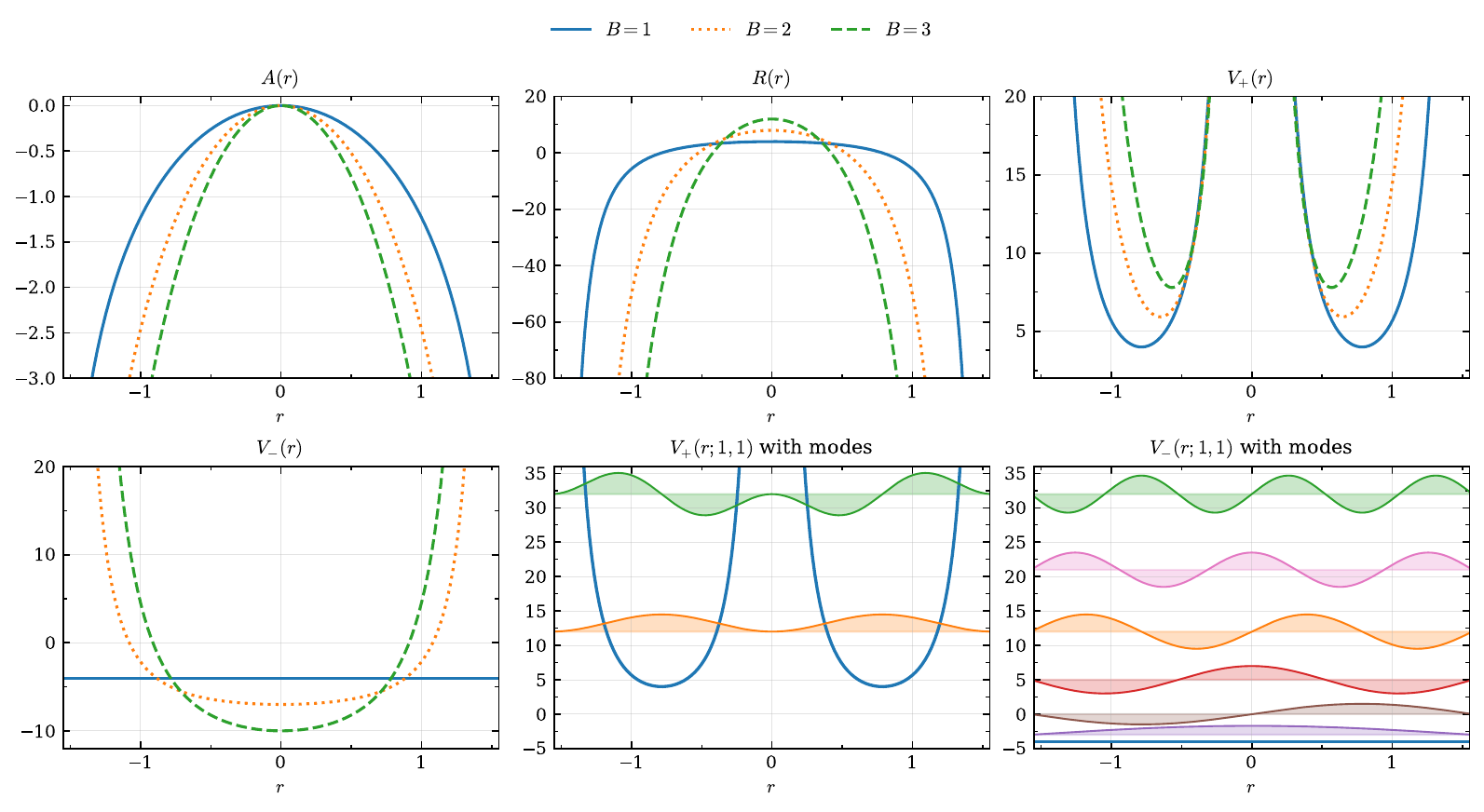}	
	\caption{Plots of the warp factor $A(r;B,1)$, the scalar curvature $R(r;B,1)$, the effective potential $V_{+}(r;B,1)$, its partner potential $V_{-}(r;B,1)$, the wave functions of $V_{+}(r;1,1)$ and $V_{-}(r;1,1)$, respectively. The blue solid line {{at}} the bottom of the last panel is the infinite square well potential $V_{-}(r;1,1)$.}
	\label{fig:1}%

\end{figure*}

For the second analytic case with $n=2$, the coordinate transformation is given by $z=\tan(r)$, and the background solutions are:
\begin{eqnarray}
\phi(z)&=&v\,{\rm arcsinh}(z),\\
A(z)&=&-\tfrac{1}{2}\log(1+  z^{2}).
\end{eqnarray}
Similarly, here $z$ runs from negative infinity to infinity. From Eq.~(\ref{34}), the effective potential becomes:
\begin{equation}\label{0011}
V(z)=\frac{3  z^2+2}{\left( z^3+z\right)^2}, \quad \text{for} \quad n=2.
\end{equation}
As shown in Fig.~\ref{fig:4}, the shape of $V(z)$ when $n=2$ is similar to the one with $n=1$, and is also a singular potential. As $|z|\to\infty$, one has $V(z)\sim 3/( z^4)$, and near the origin, $V(z)\sim 2/z^{2}$. Therefore, the potential does not support any normalizable bound states. Similar to the $n=1$ case, the entire spectrum is purely continuous. 

Since in both cases $V(z)$ admits only a gapless continuum with plane-wave asymptotics, it would be interesting to see if  the {{non-canonical}} twinlike model~\eqref{25} can produce spectrum with bound states. Because bound states have direct holographic meaning: they represent discrete excitations of boundary operators and diagnose the IR of the dual theory (a gapped, confining phase when the spectrum is discrete).

\subsubsection{Scalar perturbations in the {{non-canonical}} twinlike model}
In the {{non-canonical}} twinlike model (\ref{25}), the solutions take the same form as in Eqs.~(\ref{48})-(\ref{050}), and the function $U(\phi)$ is given by:
\begin{equation}
  U(\phi)=-\kappa \frac{\text{sech}^2\left(\frac{\phi }{v}\right) \left(\cosh ^{2 n}\left(\frac{\phi }{v}\right)-1\right)}{8 n}.\label{051}
\end{equation}

Using Eq.~(\ref{35}) and (\ref{40}), we obtain
\begin{eqnarray}
    f(r)&=&-\frac{4 \csc ( r) \cos ^{-\frac{n}{2}-1}( r)}{\kappa  v \sqrt{\tan ^2( r)+1}},\\
\mathcal{W}(r)&=&\frac{n}{2} \tan ( r)- \cot ( r).
\end{eqnarray}
The effective potential in Eq.~(\ref{34}) takes the following form:
\begin{eqnarray}\label{55}
  V_{\text{eff}}(r)&=&-\frac{(n+2)^2}{4}+\frac{n(n+2)}{4} \sec ^2(  r) + 2 \csc^2(r),
\end{eqnarray}
which is exactly the P\"oschl-Teller I potential:
\begin{eqnarray}
V_{+}(r;B,C) & =&-(B+C)^2+B(B+1) \sec ^2({{r}}) \nonumber\\
& +& C(C+1) \csc ^2({{r}}),
\end{eqnarray}
with $B=\frac{n}{2}$ and $C=1$, defined on the domain $r\in(-\frac{\pi}{2},\frac{\pi}{2})$ where $\cos(r)>0$~\cite{22}. It should be noted that this potential remain analytic expression for arbitrary values of $n>0$, as compared to previous cases where only the $n=1,2$ cases were analytically accessible. The corresponding partner potential of Eq.~(\ref{55}) is:
\begin{equation}\label{49}
V_{-}(r;B,C)=-(B+C)^2+B(B-1) \sec^2(r),
\end{equation}
with same $B$ and $C$ defined earlier. We draw $V_{\text{eff}}(r)$ in Eq.~(\ref{55}) and its partner potential in Eq.~(\ref{49}) for $B=1,2,3$ in Fig.~\ref{fig:1}. If supersymmetry is unbroken, the energy spectrum of the above two potentials maintain a one-to-one correspondence, except the ground state~\cite{22}. However, due to the presence of the term $\csc (r)$, the effective potential always has a singularity at the origin as shown in Fig.~\ref{fig:1}. This singularity breaks the supersymmetry and disrupts the spectral correspondence between the two potentials.

We show this by giving an example. For $B=1$ (i.e., $n=2$), the potential $V_{-}(r;1,1)=-4$, and becomes an infinite square well. For this case, the energy levels are given by $E_{l}^{-}=l^2+2l-3$ where $l$ is an integer starting from $0$. The first six energy levels and their associated bound states of the potential $V_{-}(r;1,1)$ are given by:
\begin{eqnarray}\label{049}
  E_{0}^{-}&=&-3; \quad \psi_{0}^{-}(r)\propto \cos (r),  \nonumber\\
  E_{1}^{-}&=&0; \quad \psi_{1}^{-}({{r}})\propto \sin (2 r), \nonumber\\
  E_{2}^{-}&=&5; \quad \psi_{2}^{-}({{r}})\propto \cos (3 r), \nonumber\\
  E_{3}^{-}&=&12; \quad \psi_{3}^{-}({{r}})\propto \sin (4 r), \nonumber\\
  E_{4}^{-}&=&21; \quad \psi_{4}^{-}({{r}})\propto \cos (5 r), \nonumber\\
  E_{5}^{-}&=&32; \quad \psi_{5}^{-}({{r}})\propto \sin (6 r), \nonumber
\end{eqnarray}
For the effective potential $V_{+}(r;1,1)=V_{\text{eff}}(r)$, the bound states can be calculated using the {{following}} formula~\cite{22}:
\begin{eqnarray}
\psi_0^{+}(r, n) & \propto& \mathcal{A}(r, n) \psi_1^{-}(r, n),\nonumber \\
& \propto& \mathcal{A}(r, n) \mathcal{A}^{\dagger}(r, n) \psi_0^{-}(r, n-1). \nonumber\\
\psi_1^{+}(r, n) & \propto& \mathcal{A}(r, n) \psi_2^{-}(r, n), \nonumber\\
& \propto& \mathcal{A}(r, n) \mathcal{A}^{\dagger}(r ,n) \mathcal{A}^{\dagger}(r, n-1)\nonumber\\
&\times& \psi_0^{-}(r, n-2) .\nonumber\\
& \vdots&\nonumber
\end{eqnarray}

The first two bound states and the corresponding energies are given by:
\begin{eqnarray}
   E_{0}^{+}&=&12; \quad \psi_{0}^{+}(r)\propto \sin ^2(r) \cos ^{2}(r),  \label{62}\\
   E_{1}^{+}&=&32; \quad \psi_{1}^{+}(r)\propto  -\sin (2 r) \sin (4 r) ,\label{062}
\end{eqnarray}
with $E_{l}^{+}=4l^2+16l+12$. In the above expressions, the subscript ``0'' in $E_{0}^{-}$ and $\psi_{0}^{-}$ denotes the ground state. Both potentials capture infinite many bound states, due to the infinite depth of the potential well. These bound state wave functions all vanish at the two boundaries, so that the unitarity boundary condition is satisfied. Moreover, it can be seen from Eqs.~(\ref{62}) and (\ref{062}) that there is no negative energy level in the spectrum, so our solution is stable against linear perturbations. We draw these bound state wave functions in Fig.~\ref{fig:1}, and we can see that the one-to-one correspondence between the energy levels of $V_{+}(r;1,1)$ and $V_{-}(r;1,1)$ is disrupted. For instance, it can be checked that the states $\psi_{0}^{+}$ and $\psi_{1}^{+}$ of $V_{+}(r;1,1)$ corresponds to $\psi_{3}^{-}$ and $\psi_{5}^{-}$ of $V_{-}(r;1,1)$, respectively. This is due to the presence of singularity at the origin of $V_{+}(r;1,1)$ that triggers supersymmetry breaking. 

\begin{figure*}[!ht]
  \centering 
  \includegraphics[width=\textwidth, angle=0]{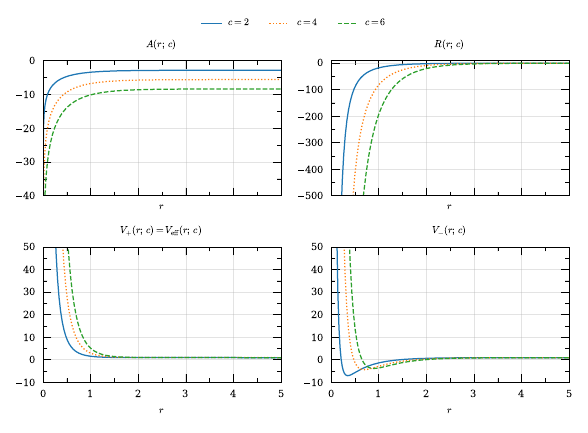}	
  \caption{Plots of the warp factor $A(r;{{c}})$, the scalar curvature $R(r;{{c}})$, the effective potential $V_{+}(r;{{c}})$ and its partner potential $V_{-}(r;{{c}})$.} 
  \label{fig:example}%
\end{figure*}

\subsection{Another solution}
\label{sec:solution2}
In this section, we study the superpotential appeared in the holographic GPPZ flow~\cite{Girardello:1999bd}:
\begin{equation}\label{03}
  W(\phi)=k v^2 \cosh \left(\frac{\phi}{v}\right)+b,
\end{equation}
in our 2D model. Note that here we add a constant $b$ to the superpotential, and it will be shown that the introduction of $b$ will lead to the Eckart potentials in the perturbation equations. Using the first-order formalism discussed previously, we obtain the following solutions:
\begin{eqnarray}
    \phi(r)&=&2 v \,{\rm arctanh }(\exp ( k r)),\label{02}\\
    A(r)&=&\frac{1}{4} \left(\kappa  v^2 \log (\sinh (k r)){{-b \kappa  r}}\right),\label{015}\\
    V(\phi)&=&{{\frac{k^2 v^2}{2}  \sinh ^2\left(\frac{\phi }{v}\right)-\frac{\kappa }{8}  \left[b+k v^2 \cosh \left(\frac{\phi }{v}\right)\right]^2.}}\label{016}
\end{eqnarray}
Similarly, the {{parameters}} $k$ and $v$ here characterize the thickness of the solution and the expectation value of the vacua, respectively. Note that the warp factor $A(r)$ is only defined for $r>0$ and divergence as $r \to 0^+$, so that our background {{space-time}} is the positive half-space. Moreover, solutions analogous to Eqs.~(\ref{02})-(\ref{016}) were analyzed in a $d+1$ scalar-gravity setup~\cite{63}. For $d=4$, the metric corresponds to the dilaton domain-wall solution of type-IIB supergravity~\cite{67,68}, first analyzed in the early AdS/CFT literature. As a type-IIB solution, however, it is non-supersymmetric.

We now consider scalar perturbations around the above background solutions. In the canonical model, the coordinate transformation $r \to z$ has no closed form, so we turn to the {{non-canonical}} model with the Lagrangian in Eq.~(\ref{25}). In this case, $U(\phi)$ becomes:  {
\begin{eqnarray}
U(\phi)&=&\frac{\text{csch} ^2(\frac{\phi}{v})}{2 v^2 k^2}\nonumber \\
  &\times& \left(1-\sinh ^{\frac{\kappa  v^2}{2}}\left(\frac{\phi }{v}\right) \tanh ^{\frac{b \kappa }{2 k}}\left(\frac{\phi }{2 v}\right)\right) \label{018}.
\end{eqnarray}
}

{{Taking $k v^2=b$,  defining $c\equiv \frac{\kappa v^2}{8}$, and using Eq.~(\ref{35}) we obtain:
\begin{equation}
f(r)=-\frac{v }{2 c}e^{(c+1) k r} \sinh ^{-c}({kr}),
\end{equation} 
for which}}  the SUSY QM superpotential in Eq.~(\ref{40}) takes the following form:
\begin{equation}\label{07}
 {   \mathcal{W}(r)=k (c+1- c \coth (k r))}.
 \end{equation}
{For simplicity, let us take $k=1$ from now on}.

The effective potential in the Schr\"odinger-like equation becomes:
\begin{eqnarray}\label{08}
    V_{\text{eff}}(r)&=&{{c}}^2+({{c}}+1)^2-2{{c}}(1+{{c}})\coth( r)\nonumber\\
    &+&{{c}}({{c}}+1) \text{csch} ^2( r).
\end{eqnarray}
This is nothing but the Eckart potential~\cite{22}:
\begin{equation}
V_{+}(r;B,C)=B^2+\frac{C^2}{B^2} -2C \coth(r)+B(B+1) \text{csch}^2(r),
\end{equation}
{{with $B=c$ and $C=c(1+c)$}}. The partner potential $V_{-}(r;B,C)$ is:
\begin{equation}\label{09}
    V_{-}(r;B,C)=B^2+\frac{C^2}{B^2} -2C \coth(r)+B(B-1) \text{csch}^2(r).
\end{equation}
In terms of paramter ${{c}}$, this is:
\begin{equation}
    V_{-}(r;{{c}})={{c}}^2+({{c}}+1)^2-2{{c}}(1+{{c}})\coth( r)+{{c}}({{c}}-1) \text{csch} ^2( r).
\end{equation}
Note that the effective potential in Eq.~(\ref{08}) is a monotonic decreasing function for any ${{c}}$ (see Fig.~\ref{fig:example} for {examples}). Since there is no potential well, the effective potential has no bound states.

\section{Summary and Outlook}
\label{sec:conclusion}
In this work we analyzed two brane-world-type solutions in the 2D MMSS model with singular space-time backgrounds. By first choosing a suitable hyperbolic superpotential, the first-order formalism yields analytic background solutions directly analogous to Ref.~\cite{64}. In the model with a canonical scalar matter field, { the linear stability potential is a singular P\"oschl-Teller~II potential which has no bound states. While for the model with non-canonical scalar, the stability potential becomes a P\"oschl-Teller~I potential with an infinite}, fully discrete tower of normalizable modes. Then inspired by the works in Refs.~\cite{Girardello:1999bd,63}, we also construct a second solution (with a different hyperbolic superpotential) whose {stability potential is} an Eckart potential without bound states. { The positivity of the linear spectra ensures the stability of the background solutions.}

The exact solvability of the perturbation equations are important for several reasons. First, the shape-invariant potentials that arise in these models are well-studied in quantum mechanics, and their exact solvability allows for a detailed analysis of the perturbation spectra. The spectra are crucial for understanding many physical properties of the 2D brane solutions. These solutions are often kink like, and the spectra determine the dynamics of kink--kink collisions~\cite{Christov:2018ecz,Manton:2018deu,Adam:2019xuc,Zhong:2019fub,Bazeia:2023qpf,Bazeia:2025yok}. Moreover, the spectra also guide the quantization of these kink solutions~\cite{Evslin:2021nsi,Evslin:2021gxs,Evslin:2022fzf,Evslin:2022clv,Ogundipe:2024ibv,Ogundipe:2024chv,Evslin:2024czr}, with discrete modes determining one-loop mass shifts and continuum phases fixing scattering data. Thus, exact solvability provides a solid foundation for both classical and quantum analyses of these models.

Moreover, the study of 2D brane world models {{with shape-invariant stability potentials provides}} an interesting framework for exploring the interplay between gravity and quantum mechanics. The presence of singularities in the background space-time raises intriguing questions about the nature of these solutions and their implications for the underlying physics. In this regard, we hope our work could provide some insights towards better understanding connections between quantum theory and gravitational models.

\begin{acknowledgements}
  This work was supported by the National Natural Science Foundation of China (Grant number 12175169).
\end{acknowledgements}



\end{document}